\documentclass[aps,showpacs,preprintnumbers,amsmath, amssymb]{revtex4}

\usepackage{float}
\usepackage{graphics,epsfig}
\usepackage{graphicx}
\usepackage{dcolumn}
\usepackage{bm}

\begin{document}
\baselineskip=0.8 cm

\title{{\bf Upper bounds on the compactness at the innermost light ring of anisotropic horizonless spheres }}
\author{Yan Peng$^{1}$\footnote{yanpengphy@163.com}}
\affiliation{\\$^{1}$ School of Mathematical Sciences, Qufu Normal University, Qufu, Shandong 273165, China}

\vspace*{0.2cm}
\begin{abstract}
\baselineskip=0.6 cm
\begin{center}
{\bf Abstract}
\end{center}

In the background of isotropic horizonless spheres,
Hod recently provided an analytical proof of a bound
on the compactness at the innermost light ring
with the dominant energy
and non-negative trace conditions.
In this work, we extend the discussion of isotropic
spheres to anisotropic spheres. With the same dominant energy
and non-negative trace conditions,
we prove that Hod's bound also holds in the case of
anisotropic horizonless spheres.

\end{abstract}

\pacs{11.25.Tq, 04.70.Bw, 74.20.-z}\maketitle
\newpage
\vspace*{0.2cm}

\section{Introduction}

According to general relativity, highly curved spacetimes
may possess closed light rings (null circular geodesics),
on which massless particles can orbit in a circle \cite{c1,c2}.
It is well known that closed light rings are usually related
to black hole spacetimes. In fact, closed light rings may also exist in the
horizonless ultra-compact spacetime.
From theoretical and astrophysical aspects, the light rings
have been extensively studied in various gravity backgrounds
\cite{ad1,ad2,ad3,ad4,ad5,c3,c4,c5,z1,z2,z3}.

The closed light ring plays an important role in understanding properties
of curved spacetimes. For example, the interesting phenomenon
of strong gravitational lensing in highly curved spacetimes
is closely related to the existence of light rings \cite{c6}.
In addition, the light ring can be used to describe
the distribution of exterior matter fields outside black holes
\cite{s1,s2,s3,s4,s5,s6}. And it was also proved that the
innermost light ring provides the fastest way to circle a central black
hole as measured by observers at the infinity \cite{s7,s8,s9}.
Moreover, the existence of stable light rings suggests that
the central compact stars may suffer from
nonlinear instabilities \cite{us1,us2,us3,us4,us5,us6,us7}.
And unstable light rings can be used to
determine the characteristic resonances of black holes
\cite{r1,r2,r3,r4,r5,r6,r7,r8}.

Recently, the compactness at the innermost closed light ring
was investigated. The compactness can be described by
the parameter $\frac{m(r)}{r}$,
where $m(r)$ is the gravitational mass within the radius r.
In the case of black holes, the compactness parameter
at the innermost light ring is characterized by the lower bound
$\frac{m(r_{\gamma}^{in})}{r_{\gamma}^{in})}\geqslant \frac{1}{3}$
with $r_{\gamma}^{in}$ as the innermost light ring radius.
However, very differently in the horizonless case, numerical data in \cite{NM1} suggests
that the compactness parameter may satisfy an upper bound
$\frac{m(r_{\gamma}^{in})}{r_{\gamma}^{in})}\leqslant \frac{1}{3}$
for spherically symmetric ultra-compact isotropic spheres.
Hod has provided compact analytical
proofs of the characteristic intriguing bound
$\frac{m(r_{\gamma}^{in})}{r_{\gamma}^{in})}\leqslant \frac{1}{3}$
for the spherically symmetric spatially regular spheres
with isotropic tensor ($p=p_{\tau}$),
where $p$ and $p_{\tau}$ are interpreted as
the radial pressure and the tangential pressure respectively \cite{AP1}.

In the present paper, we study
the compactness at the innermost light ring of
horizonless spheres. We shall prove the bound
$\frac{m(r_{\gamma}^{in})}{r_{\gamma}^{in})}\leqslant \frac{1}{3}$
in the case of anisotropic sphere with $p\neq p_{\tau}$.
We point out that this bound in the isotropic case of $p=p_{\tau}$ has been proved in \cite{AP1}.
Our main results are included in the last section.

\section{Investigations on the compactness at the innermost light ring}

We study the closed light ring of spherically symmetric
configurations. In the standard Schwarzschild coordinate,
these spacetimes are expressed by line element \cite{AP1}
\begin{eqnarray}\label{AdSBH}
ds^{2}&=&-e^{-2\delta}\mu dt^{2}+\mu^{-1}dr^{2}+r^{2}(d\theta^2+sin^{2}\theta d\phi^{2}),
\end{eqnarray}
where the metric has two functions
$\delta(r)$ and $\mu(r)=1-\frac{2m(r)}{r}$. For horizonless asymptotically flat spacetimes,
the metric functions are characterized by the near origin behavior \cite{AP1,b1,b2}
\begin{eqnarray}\label{AdSBH}
\mu(r\rightarrow 0)=1+O(r^2)~~~~~~and~~~~~~\delta(0)<\infty
\end{eqnarray}
and the far region behavior \cite{AP1,b1,b2}
\begin{eqnarray}\label{AdSBH}
\mu(r\rightarrow \infty)=1~~~~~~and~~~~~~\delta(r\rightarrow \infty)=0.
\end{eqnarray}

We denote the components of the energy-momentum tensor as
\begin{eqnarray}\label{AdSBH}
\rho=-T^{t}_{t},~~~~p=T^{r}_{r}~~~~and~~~~p_{\tau}=T^{\theta}_{\theta}=T^{\phi}_{\phi},
\end{eqnarray}
where $\rho$, $p$ and $p_{\tau}$ are respectively the energy density, the radial pressure and
the tangential pressure of the horizonless configurations \cite{s3,t1}.
For the case of isotropic energy-momentum tensor, there is the relation $p=p_{\tau}$  \cite{AP1}.
In this work, our discussion also covers the case of $p\neq p_{\tau}$.
The unknown metric functions are determined by the Einstein equations
$G^{\mu}_{\nu}=8\pi T^{\mu}_{\nu}$. With the energy density and pressures (4),
one can express the Einstein field equations in the form
\begin{eqnarray}\label{BHg}
\mu'=-8\pi r \rho+\frac{1-\mu}{r},
\end{eqnarray}
\begin{eqnarray}\label{BHg}
\delta'=\frac{-4\pi r (\rho+p)}{\mu}.
\end{eqnarray}

Using the Einstein field equations (5) and (6), it has been
explicitly proved that the closed light rings are characterized
by the relation
\cite{b1}
\begin{eqnarray}\label{BHg}
\mathcal{R}(r)=3\mu(r)-1-8\pi r^2p(r)=0~~~~for~~~~r=r_{\gamma},
\end{eqnarray}
where $r_{\gamma}$ is the radius of the closed light ring.

From equation (2) and the regular condition $p(0)<\infty$,
the function $\mathcal{R}(r)$ is characterized by
asymptotical behaviors
\begin{eqnarray}\label{BHg}
\mathcal{R}(r)=3\mu(r)-1-8\pi r^2p(r)\rightarrow 2~~~~for~~~~r\rightarrow 0.
\end{eqnarray}
We label $r_{\gamma}^{in}$ as the radius of
the innermost closed light ring, which corresponds to
the smallest positive root of $\mathcal{R}(r)=0$. In the range $[0,r_{\gamma}^{in}]$,
the function $\mathcal{R}(r)$ satisfies the relation
\begin{eqnarray}\label{BHg}
\mathcal{R}(r)\geqslant0~~~~for~~~~r\in [0,r_{\gamma}^{in}].
\end{eqnarray}
In particular, at the innermost closed light ring,
there is the relation
\begin{eqnarray}\label{BHg}
\mathcal{R}(r)=3\mu(r)-1-8\pi r^2p(r)=0~~~~for~~~~r=r_{\gamma}^{in}.
\end{eqnarray}

Substituting equations (5) and (6) into the conservation equation $T^{\mu}_{r;\mu}=0$,
one obtains a relation
\begin{eqnarray}\label{BHg}
P'(r)=\frac{r}{2g}[\mathcal{R}(\rho+p)+2gT],
\end{eqnarray}
where $P(r)=r^4p(r)$, $\mathcal{R}=3\mu-1-8\pi r^2p$ and
$T=-\rho+p+2p_{\tau}$.

In proving the following bound (22), Hod has imposed the
conditions of the dominant energy and the
non-negative trace of the energy-momentum tensor \cite{AP1}.
In this work, we also impose the same energy condition.
The dominant energy condition is
\begin{eqnarray}\label{BHg}
\rho\geqslant |p|,~|p_{\tau}|\geqslant 0.
\end{eqnarray}
And the non-negative trace condition is \cite{NM1,AP1,b1,b2}
\begin{eqnarray}\label{BHg}
T=-\rho+p+2p_{\tau}\geqslant 0.
\end{eqnarray}

In fact, for a polytropic pressure density equation of the form
$p=p_{\tau}=k_{p}\rho$, Hod has obtained a bound
$\frac{m(r_{\gamma}^{in})}{r_{\gamma}^{in})}\leqslant \frac{k_{p}+2}{6(k_{p}+1)}$,
where $\rho$, $p$ and $p_{\tau}$ are interpreted as
the energy density, the radial pressure and
the tangential pressure respectively \cite{AP1}.
In the case of $k_{p}\geqslant \frac{1}{3}$ or a non-negative trace
$T=-\rho+p+2p_{\tau}=-\rho+3p\geqslant 0$,
Hod's bound is $\frac{m(r_{\gamma}^{in})}{r_{\gamma}^{in})}\leqslant \frac{1}{3}$,
which is the same as (22).
So Hod proved the bound (22) in the case of $p=p_{\tau}$.
In the present work, we generalize the discussion to
cover the case of $p\neq p_{\tau}$.

Relations (9-13) yield that the function $P(r)$ satisfies the inequality
\begin{eqnarray}\label{BHg}
P'(r)\geqslant 0~~~~for~~~~r\in [0,r_{\gamma}^{in}].
\end{eqnarray}

Near the origin, the pressure function $P(r)$ has the asymptotical behavior
\begin{eqnarray}\label{BHg}
P(r\rightarrow 0)=0.
\end{eqnarray}

With relations (14) and (15), one obtains
\begin{eqnarray}\label{BHg}
P(r)\geqslant 0~~~~for~~~~r\in [0,r_{\gamma}^{in}].
\end{eqnarray}

The relation (16) and $P(r)=r^4p(r)$ yield that
\begin{eqnarray}\label{BHg}
p(r)\geqslant 0~~~~for~~~~r\in [0,r_{\gamma}^{in}].
\end{eqnarray}

In particular, at the innermost light ring, there is the relation
\begin{eqnarray}\label{BHg}
p(r_{\gamma}^{in})\geqslant 0.
\end{eqnarray}

According to (10) and (18), one finds that
\begin{eqnarray}\label{BHg}
3\mu(r)-1= 8\pi r^2p(r)\geqslant 0~~~~for~~~~r=_{\gamma}^{in}.
\end{eqnarray}

The relation (19) yields that
\begin{eqnarray}\label{BHg}
\mu(r_{\gamma}^{in})\geqslant \frac{1}{3}.
\end{eqnarray}

The inequality (20) can be expressed as
\begin{eqnarray}\label{BHg}
1-\frac{2m(r_{\gamma}^{in})}{r_{\gamma}^{in}}\geqslant \frac{1}{3}.
\end{eqnarray}

Then we obtain an upper bound on the compactness at the innermost light ring
\begin{eqnarray}\label{BHg}
\frac{m(r_{\gamma}^{in})}{r_{\gamma}^{in}}\leqslant \frac{1}{3}.
\end{eqnarray}

\section{Conclusions}

We studied the compactness at the innermost light ring
of anisotropic horizonless spheres. We assumed the
dominant energy and non-negative trace conditions.
At the innermost light ring, we obtained an upper
bound on the compactness expressed as
$\frac{m(r_{\gamma}^{in})}{r_{\gamma}^{in}}\leqslant \frac{1}{3}$,
where $r_{\gamma}^{in}$ is the innermost light ring
and $m(r_{\gamma}^{in})$ is the gravitational mass within the sphere of radius
$r_{\gamma}^{in}$. In fact, Hod firstly proved this bound in the spacetime of
horizonless spheres with isotropic tensor $p=p_{\tau}$ \cite{AP1}. In the present work, we
proved the same bound in the background of horizonless spheres with
generalized anisotropic tensor covering the case of $p\neq p_{\tau}$.

\begin{acknowledgments}

This work was supported by the Shandong Provincial
Natural Science Foundation of China under Grant
No. ZR2018QA008. This work was also supported by
a grant from Qufu Normal University of China under
Grant No. xkjjc201906.

\end{acknowledgments}

\end{document}